\documentclass[aps,pra,preprint,showpacs]{revtex4}
\usepackage{amsmath,graphicx}
\begin{document}
\def\la{{\langle}}
\def\ra{{\rangle}}

\title{Path summation and quantum measurements }
\author{D.Sokolovski and R. Sala Mayato$^\dagger$}
\address{School of Mathematics and Physics,
         Queen's University of Belfast, 
	 Belfast, BT7 1NN, United Kingdom,}

\address{$^\dagger$ Departamento de Fisica Fundamental II,
         Universidad de La Laguna,
	          La Laguna (S/C de Tenerife), Spain}
\date{\today}
%
%\maketitle
%+++++++++++++++++++++++++++++++++++++++++++++++++++++++++++++++++++++++++++++
%
%              Abstract
%
%+++++++++++++++++++++++++++++++++++++++++++++++++++++++++++++++++++++++++++++

\begin{abstract}
We propose a general theoretical approach to quantum measurements
based on the path (histories) summation technique.
For a given dynamical variable A,
the Schr\"odinger state of a system in a Hilbert
space of arbitrary dimensionality is decomposed into a set
of substates, each of which corresponds to a particular 
detailed history of 
the system. 
The coherence between the substates may then be destroyed by 
meter(s) to a degree determined by the nature and the accuracy
of the measurement(s) which may be
of von Neumann, finite-time or continuous type.
Transformations between the histories obtained for non-commuting 
variables and construction of simultaneous histories for non-commuting
observables are discussed.
Important cases of a particle described by Feynman paths in the 
coordinate space and a qubit in a two dimensional
Hilbert space are studied in some detail.

\end{abstract}

\maketitle
%
%}
%\documentstyle[twocolumn,aps,epsf]{revtex}
%\narrowtext
%\twocolumn
%
\section{Introduction}

Path integrals and, more generally, the path summation techniques
\cite{Feyn,vonN,Shul,Klein}
have found broad application in quantum mechanics.
One advantage of such techniques is that they reduce the task 
of calculating quantum mechanical amplitudes to summation
over certain subsets of particles histories.
As such, they provide a convenient tool for the quantum measurement
theory, where the knowledge of the system's past is equivalent
to restricting its evolution to a reduced number of scenarios.
Such restriction is usually effected by a measurement device
(meter), or an environment, with which the systems interacts
during its evolution. Thus, destruction of coherence between
the system's pasts is synonymous with a dynamical interaction,
and the two should be considered together.
An analysis of a quantum mechanical quantity based exclusively on devising
a meter for its measurement is usually incomplete, as it provides
only a limited theoretical insight into the nature of the
measured quantity \cite{MET1,RAF}.
Equally, an analysis purely in terms of 
quantum histories, such as Feynman paths \cite{PT1,PT2},
has the disadvantage of leaving open the question of how, if at all, 
the obtained amplitudes can be observed.
There are also different types of quantum measurements
to be considered: (quasi)instantaneous von Neumann
measurements \cite{vonN}, most commonly used in applications
such as quantum information theory, finite time measurements
\cite{Per} studied in \cite{S1,S2,S3,S4,S5,S6,S7,S8,S9,S10,S11,
S12,S13,S14,SBOOK} in connection
with the tunnelling time problem and continuous measurements
\cite {MBOOK,M1,M2,M3,M4,M5,M6,M7},
where a record of particle's evolution is produced by a 
'measuring medium'.
In addition, measurements of the same type differ in accuracy,
depending on the strength of interaction between the system an
a meter or an environment. Some peculiar properties of inaccurate
'weak' measurements, proposed in \cite{Ah1}, are discussed in
\cite{Ah1,AhBOOK,Ah2,SW1,SW2}.
\newline
The purpose of this paper is to suggest a general framework,
based on the path summation approach, which would describe, 
within one formalism different types of quantum measurements
of various strengths and accuracies.
The paper is organised as follows:
in Sect.2 we apply the approach of \cite{S14}
and introduce a functional differential equation
to generate a decomposition of the Schr\"odinger
state of the system corresponding the most detailed set of histories for a
particular variable $A$.
In Sect.3 we establish the link the histories obtained and
the measurement amplitudes for various meters employed to 
measure $A$.
In Sect.4 we introduce less informative coarse grained amplitudes,
taking into account finite accuracy of a meter, as well as 
a particular type of unitary transformations for the measurement
amplitudes.
In Sect.5 we show that only the paths taking the values among 
the eigenvalues of $\hat{A}$ contribute to the fine grained
amplitude introduced in Sect.2, and obtain the standard path
representations for the quantum mechanical propagator.       
In Sect.6 we show that a particular type of coarse graining 
corresponds to the continuous measurements studied in 
\cite {MBOOK,M1,M2,M3,M4,M5,M6,M7}.
In Sect.7 we consider transformation between the sets of histories
for two, possibly non-commuting, variables $A$ and $B$.
In Sect.8 we consider the special case $\hat{B}=F(\hat{A})$,
and derive the Feynman path integral representation for the
measurement amplitude, used as a starting point for the analysis
of Refs. \cite{S1,S2,S3,S4,S5,S6,S7,S8,S9,S10,S11,S12,S13,S14}.
In Section 9 we briefly discuss construction and some properties
of simultaneous histories for two non-communing variables.
Section 10 contains our conclusions.

\section{ The quantum 'recorder' equation.}

To define a particular type of observable quantum histories  
 we will
follow Ref.\cite{S14} in suggesting that distinguishing
between the pasts
 of a simple quantum system requires decomposing its current
Schr\"odinger
state $|\Psi(t)\ra$ into a set of (generally, non-orthogonal) substates 
$|\Phi[n]\ra$, where the index $n$ labels a particular history. 

\begin{figure}[h]
\includegraphics[width=12cm]{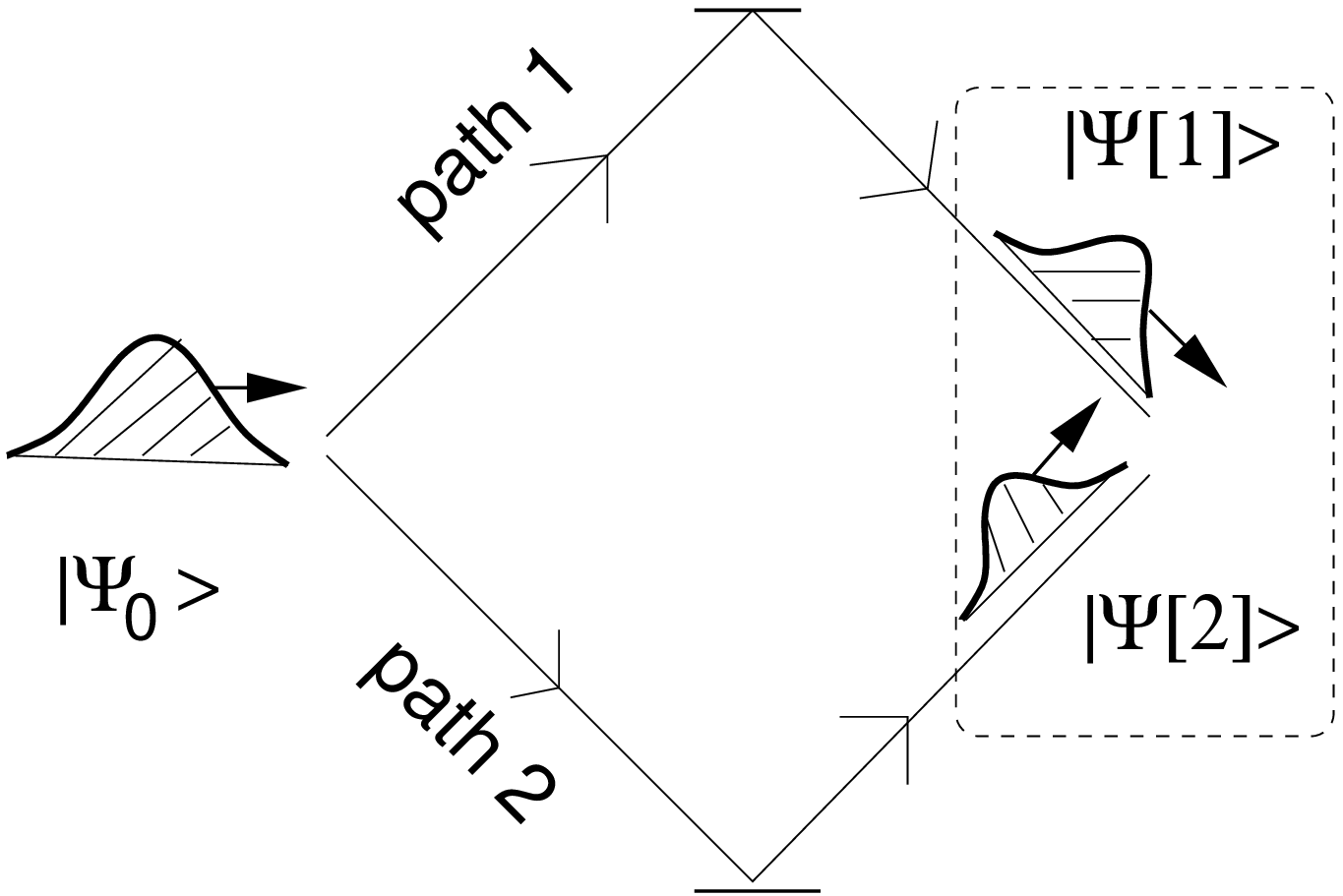}
\caption[]{
A wavepacket is split into two components,
which are later recombined.
Two substates, $|\Psi[I]\ra$ and $|\Psi[II]\ra$
correspond to the two possible histories (paths)
$I$ and $II$.
}
\label{fig:FIG0}
\end{figure}

This can be illustrated by a simple example,
equivalent to the usual two-slit experiment \cite{Feyn2}.
Let a wavepacket $|\Psi_0\ra$ be split (e.g., by means of a
beam splitter) into two parts, $|\Phi[n]\ra$, $n=1,2$
which thereafter travel along two different routes (Fig.\ref{fig:FIG0}). 
At a later time $t$, the two parts are brought together
in the same spatial region, so that the state of the system 
is a sum of two components,
\begin{equation} \label{unfold}
|\Psi(t)\ra=\sum_{n}|\Phi[n]\ra
\end{equation}
each corresponding to a particular history.
Two cases must then be considered separately.
For an isolated system in a pure state $|\Psi(t)\ra$, 
the routes are interfering alternatives,
and all information about the path travelled by the particle
is lost through quantum interference.
If, on the other hand, the two alternatives have been made,
e.g., by reversing the direction of the particle's spin when
travelling along one of the routes,
one finds the system (after tracing out the spin variable) in a mixed state.  
Observing the direction of the spin in a number of identical
trials will then show  
that the $n$-th route is travelled  with the probability
\begin{equation} \label{decoh}
W_{n}=\la\Phi[n]|\Phi[n]\ra/\sum_{m}\la\Phi[m]|\Phi[m]\ra,
\end{equation}
where $\la\psi_1|\psi_2\ra$ is the scalar product in the Hilbert space
of the particle.
\newline
Next we will use the same reasoning to study a more general
question: 
for a quantum system in the state $|\Psi(t)\ra$,
 what if anything,
can be said about the value $\varphi(t')$,
of a variable $A$, represented 
by a Hermitian operator $\hat{A}$,
at some time $t'$
within the interval
$0\le t' \le T$?
{\it A priori} it can only be assumed that $A$ may take some real
values, so that 
the set of possible histories, or paths, is that of all continuous, but not
necessary differentiable, real functions $\{\varphi(t')\}$ taking arbitrary
values at the endpoints $t=0$ and $t=T$. 
Accordingly, if $|\Phi(t|[\varphi])\ra$, yet to be defined,
is the contribution from the history $\varphi (t')$ at some $t>T$,
we should be able to obtain $|\Psi(t)\ra$ as in Eq.(\ref{unfold}),
with the sum over discrete routes replaced by functional
integration over all histories $\{\varphi(t')\}$,
\begin{equation} \label{func}
|\Psi(t)\ra=\int D\varphi |\Phi(t|[\varphi])\ra,
\end{equation}
where the symbol $D\varphi(t)$ incorporates integrations over 
over all $\varphi(t')$, including the 
endvalues $\varphi(0)$ and $\varphi(T)$ (Appendix A),
and the square brackets denote functional dependence on $\varphi$.

We define $|\Phi(t|[\varphi])\ra$ by requiring that it 
satisfies the functional differential
equation:
\begin{equation} \label{rec}
i\partial_{t}|\Phi(t|[\varphi])\ra=\{\hat{H}-i\hat{A}\frac{\delta}{\delta
\varphi(t_{-})}\}|\Phi(t|[\varphi])\ra,
\end{equation}
with the initial condition
\begin{equation} \label{init}
|\Phi(t=0|[\varphi])\ra =|\Psi_0\ra\delta[\varphi]
\end{equation}
where $|\Psi_0\ra\equiv|\Psi(t=0)\ra$ is the initial state of the system 
at t=0,  
the subscript '-' (to be omitted in the following)
indicates that the variational derivative
is taken at the time just preceding the 
current time $t$, 
and $\delta[\varphi]$ is the $\delta$-functional such that
for any functional $F[\varphi]$, the integral
$\int D\varphi F[\varphi]\delta[\varphi]=F[\varphi\equiv 0]$ (see Appendix 1).
Summing Eqs.(\ref{rec}) and (\ref{init}) over all paths $\varphi $
and using the identity 
$\int D\varphi \delta F[\varphi]/\delta\varphi(t)=0$ 
(see Eq.(\ref{A5}) of Appendix A),
shows that at any $t$, the substates $|\Phi(t|[\varphi])\ra$
add up to $|\Phi(t)\ra$, as prescribed by Eq.(\ref{func}).

\begin{figure}[h]
\includegraphics[width=10cm]{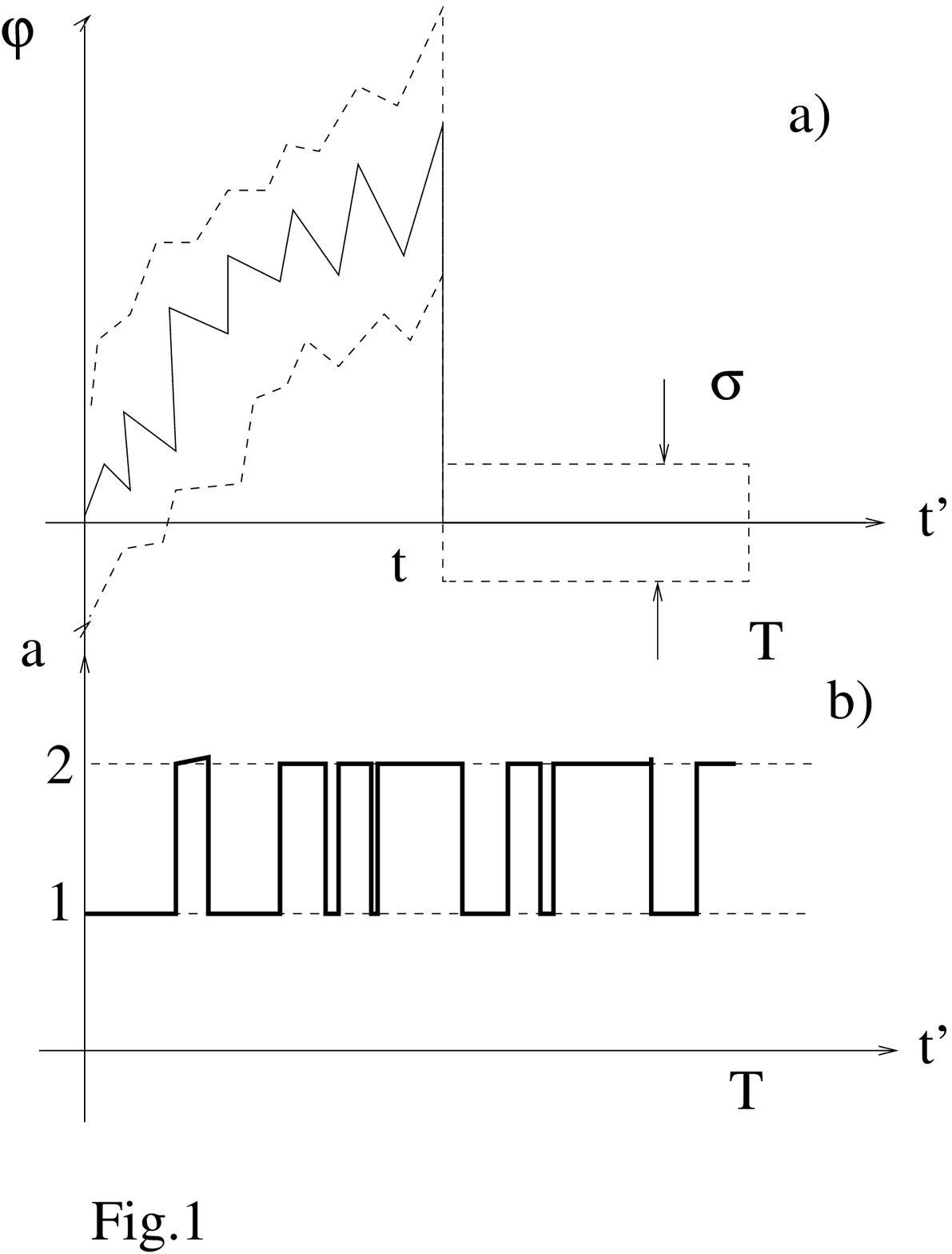}
%\vskip0.5cm
\caption[]{a)
A schematic diagramm of a path $\varphi (t')$
which contributes to $|\Phi(t|[\varphi])\ra$
at $t < T$ (solid). Also shown by a dashed line
is the tube which contains the paths, contributing
to $|\Psi(t|[\varphi])\ra$, obtained by Gaussian coarse
graining with the width $\sigma$.

b) An eigenpath which contributes to $|\Phi(T|[\varphi])\ra$
for a two-level system ( $a_1 = 1$, $a_2=2$)
}
\label{fig:FIG1}
\end{figure}

By construction, Eq.(\ref{rec})
 generates probability amplitudes for all possible histories. 
For example,
\begin{equation} \label{ampl}
A[\varphi]=\la q|\Phi(T|[\varphi])\ra
\end{equation}
yields the probability amplitude that the system, starting in 
the state $|\Psi_0\ra$ at t=0 and reaching the state  $|q\ra$ at $t=T$, has  
the history $\varphi(t')$ in the interim. 
\newline
Explicit form of $|\Psi(t|[\varphi])\ra$
can be obtained by writing it as a Fourier integral
\begin{equation} \label{Four}
|\Phi(t|[\varphi])\ra = \int D\lambda \exp[i\int_0^T\lambda (t')\varphi
(t') dt'] |\Phi(t|[\lambda ])\ra.
\end{equation}
Inserting (\ref{Four}) into (\ref{rec}) shows that the 
functional Fourier transform $|\Phi(t|[\lambda])\ra$ satisfies
the Schr\"odinger equation
\begin{equation} \label{F1}
i\partial_{t}|\Phi(t|[\lambda])\ra=\{\hat{H}+\lambda(t)\hat{A}
\}|\Phi(t|[\lambda])\ra,
\end{equation}
\begin{equation} \label{init1}
|\Phi(t=0|[\lambda])\ra=|\Psi_0\ra
\end{equation}
and, therefore, can formally be written as
$\exp[-i\int_0^t(\hat{H}+\lambda(t')\hat{A})dt']| \Psi_0\ra$.
We, therefore, have
\begin{equation} \label{F1a}
i\partial_{t}|\Phi(t|[\varphi])\ra=
\int D\lambda \exp[i\int_0^T\lambda (t)\varphi
(t) dt]
\exp[-i\int_0^T(\hat{H}+\lambda(t')\hat{A})\theta _t(t')dt']
|\Psi_0\ra,
\end{equation}
where $\theta _t (z)\equiv 1$ for $z<t$ and $0$ otherwise.
It is readily seen that by  the time $t < T$ 
the operator term only affects $t'\le t$ so that
only the histories with such that 
$\varphi(t')\equiv 0$, $t<t'<T$ may have non-zero amplitudes 
$\la q|\Phi(t|[\varphi])\ra$ (Fig.\ref{fig:FIG1}a).
This suggests the following tentative interpretation for the 'quantum recorder'
equation (\ref{rec}) and the initial condition (\ref{init}). Consider a continuous 
array of meters
with pointer positions $\varphi(t')$, $0<t'<T$ such that the meter
with the position $\varphi(t)$ 'fires' at the time $t$.
Initially, all pointers are set to zero. By a time $t<T$
some of the meters have fired, 'recording' a history $\varphi(t')$, $0<t'<t$,
while those with $\varphi(t')$, $t<t'<T$ have not yet been enacted.
Once the elapsed time has exceeded $T$, the amplitudes for all the histories
are fixed and no longer change with $t$.
The term 'quantum recorder equation' is suggested by the
analogy with a classical data recorder
monitoring the value of some variable $A$.
Note, however, that whereas in the classical case a unique record is
produced as the time progresses, the 'quantum recorder' equation
(\ref{rec}) employs the complete set $\{\varphi\}$ of all virtual 
histories and assigns a time dependent (possibly zero) substate 
$|\Phi(t|[\varphi])\ra$ to each one of them. 
This allows us to treat $\varphi(t')$ as a time-independent label,
thereby simplifying the analysis of the following Section,
where we will relate $|\Phi(t|[\varphi])\ra$ to observable
measurement probabilities.

\section{ Restricted path sums and meters}

Next we show how
some of the detailed information about the variable $A$ 
contained in the decomposition  (\ref{rec})
can be extracted by coupling the system
to a set of specially
designed meters. 
We start by demonstrating that, for $t\le T$, the integral
\begin{equation} \label{phil}
|\Phi(t|\vec{\lambda})\ra_{\beta}\equiv \int D\varphi |\Phi(t|[\varphi])\ra
\exp[i\sum_{j=1}^{M}\lambda_j\int_{0}^{T}\beta_j(t')\varphi(t')dt']
\end{equation}
where $\beta_j (t)$, $j=1,2...M$ are some known functions of time,
satisfies a Schroedinger-like differential equation (we will
omit the subscript $\beta$)
\begin{equation} \label{clockl}
i\partial_{t}|\Phi(t|\vec{\lambda})\ra=\{\hat{H}
+\sum_{i=1}^{M}\lambda_i\beta_i(t)\hat{A}\}|\Phi(t|\vec{\lambda})\ra
\end{equation}
with the initial condition
\begin{equation} \label{initl}
|\Phi(t=0|\vec{\lambda})\ra=|\Psi_0\ra.
\end{equation}
Equation (\ref{clockl})
is readily obtained if Eq.(\ref{rec}) is multiplied 
by $\exp[i\sum_{i=1}^{M}\lambda_i\int_{0}^{T}\beta_i\varphi dt']$,
integrated over $\int D\varphi$ and the term, containing the 
variational derivative $\delta/\delta\varphi(t)$, is integrated
by parts. Equation (\ref{initl}) then follows upon inserting 
Eq.(\ref{init}) into Eq.(\ref{phil}).
Taking a further Fourier transform with respect to 
$\vec{\lambda}$, ($\vec{\lambda}\vec{f}\equiv\sum_{j=1}^{M}\lambda_jf_j$)
\begin{equation} \label{Four2l}
|\Phi(t|\vec{f})\ra\equiv(2\pi)^{-M}\int_{-\infty}^{\infty}
\exp(i|\vec{\lambda}\vec{f})|\Phi(t|\vec{\lambda})\ra d\vec{\lambda},
\end{equation}
yields
\begin{equation} \label{clockf}
i\partial_{t}|\Phi(t|\vec{f})\ra =\{\hat{H}
-i\sum_{j=1}^{M}\partial_{f_j}\beta_j(t)\hat{A}\}|\Phi(t|\vec{f})\ra
\end{equation}
\begin{equation} \label{initf}
|\Phi(t=0|\vec{f})\ra =|\Psi_0\ra \prod_{j=1}^{M}\delta(f_j).
\end{equation}
It is seen that Eq.(\ref{clockf}) describes a system 
interacting with $M$ external meters via time-dependent couplings
$-i\partial_{f_j}\beta_j(t)\hat{A}$,
which involve the the measured quantity, $\hat{A}$,
a swithching function $\beta_j(t)$ 
and the pointer's momentum, $-i\partial_{f_j}$.
The meters, whose pointer positions are $f_i$,
are initially prepared 
in the product state (\ref{initf}) and, after
tracing out the pointer variable the system is described
by the density operator 
\begin{equation} \label{rho}
\hat{\rho} = \int d\vec{f} |\Phi(t|\vec{f})\ra \la \Phi(t|\vec{f})|.
\end{equation}
Reading the meter one, therefore, obtains information about the
system's past.
\newline
The nature of the information obtained is clarified by noting that
interchanging the order of integration over $D\varphi$ and $\vec{\lambda}$
in Eqs.(\ref{phil}) and (\ref{Four2l})
yields
\begin{equation} \label{Four2la}
|\Phi(t|\vec{f})\ra =\int D\varphi\prod_{j=1}^{M}\delta
(F_j[\varphi]-f_j) |\Phi(t|[\varphi])\ra
\end{equation}
where the functionals $F_j[\varphi]$ are defined by
\begin{equation} \label{Four2lb}
F_i[\varphi]\equiv\int_{0}^{T}\beta_i(t')\varphi(t')dt'. 
\end{equation}
Thus, $|\Phi(t|\vec{f})\ra$ is given by a restricted path sum, in which the summation
is limited only to those histories, for which
$$F_j[\varphi]=f_j,\quad j=1,2,...M.$$
\newline
Thus, the fixed set of paths $\{\varphi \}$
has been divided, according to the values if the functionals, $\vec{f}$, into classes within which
the individual paths cannot be told apart.
The classes play the role of alternative 'routes' along which
the system may evolve from its initial state and a time dependent
probability amplitude can be assigned to each of them.
One can, therefore, analyse the  measurement process either in terms
of dynamical interaction with the pointer degrees
of freedom, or, which is conceptually much simpler, in terms
of converting interfering histories into exclusive ones \cite{Feyn}. 
Note that only part of the detailed information, contained in
the full path decomposition $|\Phi(t|[\varphi])\ra$ is extracted by the meters,
which employ $|\Phi(t|\vec{f})\ra$ and allow the rest of it to be lost through
the residual interference between the paths of the same class.
The most common types of such measurements are:
\newline A {\it von Neumann measurement}
for which $M=1$, $\beta_i(t)=\delta(t-t_0)$ and which 
determines the instantaneous value of an operator $\hat{A}$ at some $t=t_0$ 
\cite{vonN}.
\newline A {\it finite time measurement}, $M=1$, $\beta(t)=const$,
which determines a time average of an operator $\hat{A}$ over
the time $T$. Measurements of this type were first discussed in
\cite{Per} and extensively studied in connection with the tunnelling time
problem \cite{S1,S2,S3,S4,S5,S6,S7,S8,S9,S10,S11,S12,S13,S14,SBOOK}.
\newline A {\it continuous measurement},where  $M\rightarrow \infty$, 
$\beta_i(t) \sim \delta(t-t_i)$.
In this limit, a sequence of values $f_j$, $j=1,2,...M$
is replaced by a continuous function $ f(t)$, $\vec{f}\rightarrow f(t)$.
Continuous measurements, which model a particle in a 'measuring medium',
are analysed in \cite{MBOOK,M1,M2,M3,M4,M5,M6,M7}. 
This list is not exhaustive, and one can envisage various
sequences and combinations of von Neumann, finite time and 
continuous measurements.

\section{ Coarse graining and unitary transformations}

The scalar product  $\la\Phi(T|\vec{f})|\Phi(T|\vec{f})\ra$
cannot yet be interpreted 
as the probability to measure the values $\vec{f}$ 
because the $\delta $-function in Eq.(\ref{initf}),
$\delta(\vec{f})$, is not normalisable and should, therefore,
be replaced by some square-integrable function $G(\vec{f})$, 
representing a physical initial state of the meter.

To see how such initial states can be described in the
path summation approach, we note that the superposition principle
allow one to also consider more general histories represented 
by linear combinations,
with complex valued coefficients,
of the paths $\varphi$ (e.g., $\varphi'(t')=a \varphi_1(t')+
b\varphi_2(t')$), so that their contributions to the 
Schr\"odinger state of the system at $t$ is given by the linear
combinations of the corresponding substates (e.g., 
$|\Phi(t|[\varphi'])\ra =a|\Phi(t|[\varphi_1])\ra +b|\Phi(t|[\varphi_2])\ra$).
We note further that a solution of Eq.(\ref{F1}),
multiplied by an arbitrary functional $\tilde{G}[\lambda]$
remains a solution. Equivalently, as the convolution
property, Eq.(\ref{ACON}), demonstrates, the set of states
\begin{equation} \label{C1}
|\Psi(t|[\varphi])\ra \equiv \int D\varphi'G[\varphi-\varphi']|\Phi(t|[\varphi])\ra, 
\end{equation}
where $G[\varphi]$ is the Fourier transform of $\tilde{G}[\lambda]$,
satisfies Eq.(\ref{rec}) with the initial condition
\begin{equation} \label{C2}
|\Psi(t=0|[\varphi])\ra = \int D\varphi'G[\varphi-\varphi']\delta[\varphi']
|\Psi_0\ra=G[\varphi]|\Psi_0\ra.
\end{equation}
Repeating the argument of the previous Section shows 
that the restricted sum over histories
\begin{equation} \label{C3}
|\Psi(t|\vec{f})\ra =\int D\varphi\prod_{i=1}^{M}\delta
(F_i[\varphi]-f_i) |\Psi(t|[\varphi])\ra
\end{equation}
is the solution of the meter equation (\ref{clockf}) with the initial 
condition
\begin{equation} \label{C4}
|\Psi(t=0|\vec{f})\ra = \int D\varphi \prod_{i=1}^{M}\delta
(F_i[\varphi]-f_i)G[\varphi]
|\Psi_0\ra\equiv G(\vec{f})|\Psi_0\ra.
\end{equation}
Thus, choosing the functional $G$ in Eq.(\ref{C2})
to be 
\begin{equation} \label{C4a}
G[\varphi]=G(F_1[\varphi],F_2[\varphi]...F_M[\varphi])
\end{equation}
yields the solution of Eq.(\ref{clockf}) with the initial condition
\begin{equation} \label{C5a}
|\Psi(t=0|\vec{f})\ra = G(\vec{f})|\Psi_0\ra ,
\end{equation}
which can also be obtained by
first restricting the fine grained path sum as in Eq.(\ref{Four2l})
and then convolving the result with $G(\vec{f})$, in the $\vec{f}$-variable, \cite{SBOOK}
\begin{equation} \label{C5}
|\Psi(t|\vec{f})\ra = \int d\vec{f'} G(\vec{f}-\vec{f'})|\Phi(t|\vec{f})\ra. 
\end{equation}
The validity of Eq.(\ref{C5}) can be checked by direct substitution
into Eq.(\ref{clockf}).
The result (\ref{C5}) can be used in two different ways.

{\it 1. Coarse graining.} If $G(\vec{f})$ is chosen to be a
square-integrable function sharply peaked around $\vec{f}=0$,
e.g.,

\begin{equation} \label{C5b}
G(\vec{f})=\exp[-\sum_{i=1}^M  f_j^2/\Delta f_j^2],
\end{equation}
the coarse grained \cite{FOOTC} set $|\Psi(t|\vec{f})\ra$ corresponds to 
a measurement in which obtaining a readout $\vec{f}$ 
guarantees that in the values of the functionals $F_j$, $j=1,2,..M$
in Eq.(\ref{F1}) were $f_j$ within the error margin
$\Delta f_j$. 
By construction, 
\begin{equation} \label{C6}
W(\vec{f}) = \la\Psi(T|\vec{f})|\Psi(T|\vec{f})\ra
\end{equation}
yields the corresponding probabilities to find the pointers at positions $f_1,f_2,...f_M$ after 
the measurement
is completed at t=T. Note that this probabilities do not, in general
add to one, but can be normalised since
\begin{equation} \label{C7}
\int d\vec{f} W(\vec{f}) = \int d\vec{f} |G(\vec{f})|^2 \la\Psi_0|\Psi_0\ra 
< \infty.
\end{equation}

We have, therefore, achieved our aim of relating 
the results of measurements conducted with the help of meters,
dynamically coupled to the system,
and the possible system's histories introduced in Sect.2.
In this connection it is worth recalling 
the relation between the accuracy of a measurement 
and the strength of the coupling between the measured system
and the meter(s) \cite{SBOOK}. Indeed,
the resolution of the meters can be improved,
by replacing the initial state $G(\vec{f})$ by $G(\alpha\vec{f})$,
 $\alpha > 1$.
A change of variables $\vec{f} \rightarrow \alpha \vec{f}$ shows that
the resulting finer set of substates satisfies Eq.(\ref{clockf})
with the old initial condition, $|\Psi(t=0|\vec{f})\ra = G(\vec{f})|\Psi_0\ra$, but with the
coupling term increased $\alpha$-fold, $i\alpha \sum_{j=1}^{M}\partial_{f_j}\beta_j(t)\hat{A}$.
The same can be observed by writing $|\Psi(t|\vec{f})\ra$ as
\begin{equation} \label{C8}
|\Psi(t|\vec{f})\ra=\int G(\vec{\lambda}) \exp \{ -\int_0^t [\hat{H}+\hat{A}\sum_{i=1}^{N}\lambda_i
\beta_i(t)dt]\}|\Psi_0\ra,
\end{equation}
where $G(\vec{\lambda})$ is the Fourier transform of $G(\vec{f})$,
which shows that the substate is obtained by evolving the initial
state of the system with the Hamiltonians involving all possible magnitudes
of the coupling. Among these, only the 
$\vec{\lambda}=0$ term corresponds to the unperturbed evolutions,
while the rest contain the effects of the meter.
As the coarse graining becomes finer, $G(\vec{f})\rightarrow G(\alpha\vec{f})$,
the Fourier transform becomes broader, 
$G(\vec{\lambda})\rightarrow \alpha^{-1} G(\alpha\vec{\lambda}/\alpha)$,
and the number of $\vec{\lambda}\neq 0$ which contribute to the 
formation of the substate $|\Psi(t|\vec{f})\ra$ increases. 

{\it 2. Unitary transformations}.
The choice of $G$ in Eq.(\ref{C5})
in the form of a unitary kernel,
\begin{eqnarray} \label{C10}
G[\varphi] &\equiv& U[\varphi]\nonumber\\
\int d\vec{f''} U^*(\vec{f''}-\vec{f'})U(\vec{f''}-\vec{f})&=&\delta(\vec{f}-\vec{f'}).
\end{eqnarray}
does not provide a physical
measurement amplitude for a set of meters, but rather a unitary transformation
for the fine grained set of substates, and next we consider its physical meaning.
For the Fourier transform of $U(\vec{f})$, $U(\vec{\lambda})$, Eq.(\ref{C10})
implies $U^*(\vec{\lambda})U(\vec{\lambda})=1$, or,
\begin{equation} \label{C11}
U(\vec{\lambda})=\exp[i\eta(\vec{\lambda})],
\end{equation}
where $\eta(\lambda)$ is a  real phase.
Consider the simplest choice
\begin{equation} \label{C12}
\eta(\lambda) =-\sum_{j=1}^M a_j \lambda_j
\end{equation}
which yields
\begin{equation} \label{C13}
G(\vec{f})=\delta(\vec{f}-\vec{a})
\end{equation}
so that the transformation (\ref{C3}) corresponds 
to a shift of the zero position of the $j$-th pointer by $a_j$. 

For the phase that is quadratic in $\lambda$,
\begin{equation} \label{C14}
\eta(\lambda) =-\sum_{j=1}^M b_j \lambda_j^2
\end{equation}
we have
\begin{equation} \label{C14a}
U(\vec{f})=(2\pi)^{-M}\prod_{j=1}^N (\pi/ib_j)\exp(if_j^2/4b).
\end{equation}
Comparing the last term in Eq.(\ref{C14}) with the propagator
of the free particle with a mass $m$, \cite{Feyn},
$g(f,\tau)=(m/2\pi i \tau )^{1/2}\exp(imf^2/2\tau)$,
we note that, apart from an unimportant constant factor,
initial state of the $j$-th meter has been obtained from $\delta(f_j)$
by the free-particle evolution with $m/\tau=1/2b_j$.
Thus, the transformation (\ref{C14}) yields a fine grained amplitude
for the case when, prior to the measurement, uncoupled meters have been
allowed to evolve from their initial sharply-peaked states.
The coarse graining (\ref{C5}) and
the unitary transformation (\ref{C10}) operations commute and can 
be applied in any order, in order to produce measurement amplitudes
for different degrees of resolution and initial meter states.

\section{ Eigenpaths. Feynman path integral. Path sum for a
two-level system.}

Next we show that
Eq.(\ref{rec}) generates, 
a non-zero substate $|\Phi(t|[\varphi])>$
only for a paths such that at any given time $t'$ the value of
$\varphi(t')$ coincides with one of the eigenvalues $a_i$
of the $\hat{A}$.
Throughout this Section we will assume that
 $a_k$, are non-degenerate.
Depending on the operator $\hat{A}$, the set of such eigenpaths,
$\{a\}$ may coincide with $\{\varphi\}$ or form a smaller
subset of the latter.
Consider the time-discretised version of Eq.(\ref{F1a}),
whereby we slice the time interval $[0,T]$ into
$N$ subintervals $\epsilon$, so that 

$$t_j\equiv (j-1)\epsilon, \quad z_j\equiv z(t_j),
\quad K\equiv int\{min(t,T)\}/\epsilon$$
Thus the operator in the r.h.s. of Eq.(\ref{F1a}) takes the form
\begin{equation} \label{EG1}
\exp\{-i\int_0^t[\hat{H}+\lambda(t')\hat{A}]dt'\}
= \lim_{N\rightarrow \infty} \prod_{j=1}^{K}
\exp[-i\lambda_j \hat{A}\epsilon]
\exp[-i\hat{H}\epsilon ] 
\end{equation}
where we have made use of the Trotter product formula \cite{Shul}
(see also Appendix B)
to factorise the exponentials containing $\hat{H}$ and $\lambda \hat{A}$.
Using 
\begin{equation} \label{EG2}
\exp[-i\lambda_j \hat{A}\epsilon]=\sum_{k} \exp(-i\lambda a_k)
|a_k\ra \la a_k|
\end{equation}
and performing integrations over $\lambda_j$, $j=1,2...M-1$
yields
\begin{equation} \label{EG3}
|\Phi(t|[\varphi])\ra=\sum_{[a]}\delta[\varphi-\theta _t a]|\Phi_t[a]\ra,
\end{equation}
\begin{equation} \label{EG3a}
|\Phi_t[a]\ra \equiv \hat{U}_t[a]|\Psi_0\ra
\end{equation}
where
\begin{equation} \label{EG4}
\hat{U}[a]\equiv lim_{N\rightarrow \infty}\prod_{j=1}^{K}
|a_{k_j}\ra \la a_{k_j}|
\exp(-i\hat{H}\epsilon),
\end{equation}
$a_{k_j}\rightarrow a(t')$, and we have introduced the notation
\begin{equation} \label{EG5}
\sum_{[a]} Z[a] \equiv lim_{N\rightarrow \infty}
\sum_{k_j} Z(a_{k_1},a_{k_2},...a_{N}).
\end{equation}
Therefore, a path
$\varphi(t')$ corresponds to a non-zero substate $|\Phi(t|[\varphi])\ra$
if, and only if, at any time $t'\le t$, $\varphi(t')=a_k$, in which
case the substate itself is the eigenstate corresponding to
the eigenvalue $\varphi(t)$. It is easy to check that in Eq.(\ref{EG3})
each term in the sum over the eigenpaths $a(t')$ satisfies
Eq.(\ref{rec}) with the initial condition (\ref{init}) (see Appendix C).

For $t=T$, inserting Eq.(\ref{EG3}) into Eq.(\ref{C3})
gives the expression of the measurement amplitude as a restricted
sum over
eigenpaths,
\begin{equation} \label{EG51}
|\Phi(t|\vec{f}\ra=\sum_{[a]} \prod_{j=1}^{M}\delta
(f_j-F_j[a]) |\Phi_T(t|[a])\ra,
\end{equation}
where $F_j[a]\equiv\int_{0}^{T}\beta_j(t')a(t')dt'$.

Integrating Eq.(\ref{EG3}) over $D\varphi$ gives the path expansion of the propagator,
\begin{equation} \label{EG6}
\la a|\Psi(T)\ra = \sum_{[a']}\la a|\hat{U}_T[a']|\Psi_0\ra
\end{equation}
together with the identity
\begin{equation} \label{EG7}
\sum_{[a']}\hat{U}_T[a']|=\exp(-i\hat{H}T)
\end{equation}
The nature of the summation over $a'$ depends on the spectrum
of $\hat{A}$ and next we consider two important examples.

{\it The  Feynman path integral}.
For a one-dimensional particle of mass $m$ in a potential $V(x)$
coordinate histories are generated by the equation
\begin{equation} \label{EG8}
i\partial_{t}|\Phi(t|[\varphi])\ra =
\{-\partial_x^2/2m+V(x)-ix\frac{\delta}{\delta\varphi(t)}\}|\Phi(t|[\varphi])\ra.
\end{equation}
As the position operator $\hat{x}=x$ has a continuous spectrum extending
from $-\infty$ to $+\infty$, the set of paths $\{x(t)\}$ in Eq.(\ref{EG3})
coincides with $\{\varphi(t)\}$ in Eq.(\ref{func}),
the sum $\sum_{[a]}$ becomes $\int dx_j$, and the path sums
(\ref{EG3}) and (\ref{func}) are essentially the same.
Further, the standard derivation shows (see, for example Ref.\cite{Shul})
\begin{equation} \label{EG9}
\la x|\hat{U}_T[x]|x'\ra =lim_{N\rightarrow \infty}
(m/2\pi\epsilon)^{N/2}\exp(iS[x]),
    \end{equation}
where $S[x]=\int_0^T [m\dot{x}^2/2-V(x)]dt']$ is the classical action,
and Eq.(\ref{EG6}) becomes the familiar expression for the Feynman
propagator \cite{Feyn}.
Measurement amplitudes obtained by restricting the Feynman path integral (\ref
{EG7})
have been often studied in literature (see, for instance 
Refs.\cite{S1,S2,S3,S4,S5,S6,S7,S8,S9,S10,S11,S12,S13,S14} and \cite{MBOOK}).

{\it Path sum for a two-level system (qubit).}
Another example is a two-level system in a two-dimensional
Hilbert space. A two-dimensional version of Eq.(\ref{rec}) has the form
\begin{equation} \label{EG10}
i\partial_{t}|\Phi(t|[\varphi])\ra =\left(
\begin{array} {cc}
\epsilon_1 & V \\
V & \epsilon_2
\end{array}\right)
|\Phi\ra
-
i\frac{\delta}{\delta
\varphi(t)}
\left(
\begin{array} {cc}
1 & 0 \\
0 & 2
\end{array}\right)
|\Phi\ra
\end{equation}
where, without loss of generality, we have ascribed 'coordinates'
$1$ and $2$ to the first and second states, respectively, and
$|\Phi(t|[\varphi])\ra$ is a two-component vector in the representation
in which the 'position operator', given by the second matrix
on the right, is diagonal.
Now the eigenpaths $a(t')$ in Eq.(\ref{EG4}) can only take the 
values $1$ or $2$ at any given time, which they can change at
any $t'$ (Fig.\ref{fig:FIG1}b). Each such jump is facilitated by the
the off-diagonal part of the Hamiltonian, proportional 
to $V$.
Thus, rearranging in Eq.(\ref{EG7}) the paths according to the number of jumps
and summing over all paths
yields the expansion of the evolution operator in powers of $V$
\begin{eqnarray} \label{EG9a}
\hat{U}(T)=1+\sum_{n=1}^{\infty}(-i)^n
\int_0^T dt_n \int_0^{t_n} dt_{n-1}...\int_0^{t_2} dt_{1}\\ \nonumber
\exp[-i\hat{H}(T-t_n)]V\exp[-i\hat{H}(t_n-t_{n-1})V...V
\exp[-i\hat{H}t_1]
    \end{eqnarray}
which is the standard decomposition of the perturbation theory \cite{Mess}.
Measurement amplitudes obtained by restricting the path sum for a
two-level system have been used in Ref.\cite{S13} to analyse
the residence time problem.

\section{The Mensky's formula and continuous measurements.}

Consider next a special case of the transformation (\ref{C1}), with
\begin{equation} \label{D1}
G[\varphi]=\exp[-i\int_0^T g(t',\varphi)dt'].
\end{equation}
With the help of (\ref{EG3}) we obtain
\begin{equation} \label{D2}
|\Psi(t|[\varphi])\ra = \sum_{[a]}\exp[-ig(t',\varphi-\theta_t a)dt']
\hat{U}_t[a]|\Psi_0\ra ,
\end{equation}
where the last operator is given by the discretisation
\begin{eqnarray} \label{D3}
\exp[-ig(t',\varphi-\theta_t a)dt']\hat{U}_t[a]&=&\nonumber\\
lim_{N\rightarrow \infty}\prod_{j=1}^{N} |a_{k_j}\ra \la a_{k_j}|
\exp(-i\theta_t (t_j)\hat{H}\epsilon)\exp[-ig(t_j,\varphi-\theta_{t}(t_j) a_{k_j})\epsilon].
\end{eqnarray}
Applying the Trotter formula (\ref{T3}) to recombine
the two exponentials, and summing over the eigenpaths yields a 
compact expression for $|\Psi(t|[\varphi])\ra$,
\begin{equation} \label{D4}
|\Psi(t|[\varphi])\ra = \exp\{-i\int_0^T[\theta_{t}\hat{H}+
g(t',\varphi-\theta_t \hat{A})]dt'\}|\Psi_0\ra
\end{equation}
which for $0<t<T$ satisfies the 'recorder' equation (\ref{rec})
with the initial condition Eq.(\ref{C2}).
It follows from Eq.(\ref{D4}) that
\begin{equation} \label{D5}
|\Psi(t|[\varphi])\ra = \exp[-i\int_t^T
g(t',\varphi) dt']|\Psi_{\varphi}(t)\ra
\end{equation}
where $|\Psi_{\varphi}(t)\ra$ satisfies the effective Schr\"odinger
equation
\begin{equation} \label{D6}
i\partial_{t}|\Psi_{\varphi}(t)\ra =\{\hat{H}+g(t,\varphi-\hat{A})\}
|\Psi_{\varphi}(t)\ra,
\end{equation}
\begin{equation} \label{D7}
|\Psi_{\varphi}(0)\ra =|\Psi_0\ra .
\end{equation}
The problem of evaluating the restricted path sum for $|\Psi(t|[\varphi])\ra$ in Eq.(\ref{C1}), 
therefore,
has been reduced to solving a time-dependent Schr\"odinger equation
with the time dependence determined by $\varphi(t)$.
\newline Equations (\ref{D4}) and (\ref{D5}) were first suggested by Mensky \cite {MBOOK}
for the case when the functional $G[\varphi]$ reaches its maximum
value for $\varphi(t')\equiv 0$ and rapidly falls off as $\varphi$
deviates from zero, so that $G[\varphi]$ coarse grains $|\Phi\ra$ as
discussed in Sect.3. One such choice is
\begin{equation} \label{D8}
g(\varphi)=-i\varphi^2/\sigma^{2}
\end{equation}
which ensures that only the eigenpaths  $\varphi'$ in a tube of the 
width $\sigma$ around $\varphi$ contribute to $|\Psi(t|[\varphi])\ra$
in Eq.(\ref{C1}) (see Fig.\ref{fig:FIG1}a), and the effective Schr\"odinger equation 
Eq.(\ref{D6}) contains a non-Hermitian imaginary term 
$$-ig(t,\varphi-\theta_{t} \hat{A})]dt'|\Psi_0\ra .$$

One notes that Eq.(\ref{D4}) for $|\Psi(T|[\varphi])\ra$ can,
with the help of Eq.(\ref{A1}), also 
be written as the limit of the time-discretised form 
\begin{eqnarray} \label{D9}
|\Psi(T|[\varphi])\ra &=&\lim_{M\rightarrow \infty}
\int  \prod_{j=1}^M df'_j \exp[-(\varphi_j-f_j)^2\epsilon/\sigma^2]
\nonumber\\
\int D\varphi' \delta\bigg( f'_j&-&\int_0^T\delta(t'-t_j)\varphi'(t')dt'\bigg)
|\Phi(T|[\varphi'])\ra .
\end{eqnarray}
Comparing Eq.(\ref{D9}) with Eq.(\ref{clockf})
shows that the second integral in (\ref{D9}) is the fine 
grained amplitude for an array of $M$ von Neumann meters,
each firing at $t_j=j\epsilon$, $1\le j \le M$.
Upon integration over $df'_1, df'_2, ... df'_M$,
this amplitude is coarse grained with a product of Gaussians
whose widths increase as  $\sigma/\epsilon^{1/2}$ for 
$\epsilon=T/M \rightarrow 0$.
Thus, this is a sequence of very inaccurate
'weak' \cite{Ah1,AhBOOK,Ah2,SW1,SW2} measurement, by a set of meters
weakly coupled to the system.
Taking the limit $\sigma \rightarrow 0$ in expression (\ref{D8})
yields the solution  $|\Phi_W[\varphi)\ra$, which satisfies
the initial condition
\begin{equation} \label{D9a}
|\Phi_W(t=0|[\varphi])\ra\approx \delta(\int_0^T\varphi^2(t')dt') |\Psi_0\ra .
\end{equation}
The condition (\ref{D9a}), 
which requires that the mean-square deviation
of $\varphi$ from zero must vanish,
is similar to  Eq.(\ref{init}) which 
needs $\varphi(t')$ to vanish point-wise,
and either set of substates can be used 
for calculating the fine grained finite time measurement
amplitude (see Appendix D).
The set $|\Phi_W[\varphi]\ra$, which corresponds
to scattering by 'measuring medium', was first suggested in \cite{MBOOK}.
Equations (\ref{D6}) and (\ref{D9})
can be applied to various coarse graining and 
unitary transformations.
 No similar formulae exist, in general,
for less informative finite time measurements, which yield
information about
certain global properties of the paths, e.g., the  value
of $\int_0^T \beta(t') \varphi(t') dt'$, while 
precise values of $\varphi(t')$ remain indeterminate.
In that case,
$|\Psi(t|\vec{f})\ra$ cannot be obtained from $|\Psi_0\ra$ by evolution
with a generalised Hamiltonian, containing $\vec{f}$ as a parameter.

\section{ Transformations between representations}

Theory of representations plays an important role in quantum theory
and next we establish how the set of substates 
$|\Phi^A(T|[\varphi])\ra$, corresponding to an operator
$\hat{B}$ can be obtained 
from the set $|\Phi^B(T|[\varphi])\ra$ corresponding to another
operator in the same Hilbert space, $\hat{A}$, which may not commute with $\hat{B}$.
Defining an operator 
\begin{equation} \label{Rep1}
\hat{U}[\varphi-\varphi']\equiv
\int D\lambda \exp[i\lambda(\varphi-\varphi')]
\exp\{-i\int_0^T[\hat{H}+\lambda (t)\hat{B}]dt\}
\exp\{i\int_0^T[\hat{H}+\lambda (t)\hat{A}]dt\}
\end{equation}
and taking into account (\ref{EG3}) it is easy to show that
\begin{equation} \label{Rep2}
|\Phi^B[\varphi]\ra =\int D\varphi' \hat{U}[\varphi-\varphi']
|\Phi^A[\varphi']\ra .
\end{equation}
This expression is similar to Eq.(\ref{C1})
except that in place of the unitary kernel $U[\varphi-\varphi']$
it contains the unitary operator-valued kernel $\hat{U}[\varphi-\varphi']$,
\begin{equation} \label{Rep3}
\int D\varphi'' \hat{U}^*[\varphi''-\varphi] \hat{U}[\varphi''-\varphi']=
\delta[\varphi-\varphi'].
\end{equation}
Representing, as in Sect.5, each of the exponentials in Eq.(\ref{Rep1})
as products over infinitesimal time intervals, and applying the 
Trotter formula (\ref{T3}), we can reduce  Eq.(\ref{Rep2}) 
to
\begin{equation} \label{Rep4}
\hat{U}[\varphi-\varphi']=
\sum_{[a]}\sum_{[b]} \delta[\varphi-\varphi'-b+a]
\hat{U}_T[b(t')] \hat{U}_T^*[a(t')],
\end{equation}
where, as before, $\sum_{[z]}$ denotes the sum over all eigenpaths
corresponding to an operator $\hat{Z}$.
It is straightforward to verify that
\begin{equation} \label{Rep4a}
\sum_{[a]}\sum_{[a']}
\delta[\varphi - b +a-a']\hat{U}^{*}_T[a] \hat{U}_T[a']
=\delta[\varphi - b]\sum_{[a]}\hat{U}^{*}_T[a] \hat{U}_T[a]=\delta[\varphi - b].
\end{equation}
Inserting Eq.(\ref{EG51}) into Eq.(\ref{Rep2}) and using Eq.(\ref{Rep4a})
we  may write 

\begin{equation} \label{Rep5}
|\Phi^B[\varphi]\ra =\sum_{[a]}\delta[\varphi-b]
\hat{U}_T[b] \hat{U}_T^*[a]|\Phi^A[a]\ra
\end{equation}

and, for the coefficients in the expansions
$$ |\Phi^A[a]\ra =\sum c_{a'}|a'\ra ,
\quad
|\Phi^B[b]\ra =\sum d_{b'}|b'\ra$$
we have 
\begin{equation} \label{Rep6}
d_{b'}[b]=\sum_{a'}\sum_{[a]}
\la b'|\hat{U}_T[b] \hat{U}_T^*[a]|a'\ra c_{a'}[a].
\end{equation}
We note that for $\hat{B}\equiv \hat{A}$
Eq.(\ref{Rep1}) becomes an identity, yet
$\la b'|\hat{U}[\tilde{a}] \hat{U}^*[a]|a'\ra\ne \delta[a-\tilde{a}]$.
This is because, by construction,
the number of substates $|\Phi^A[\varphi]\ra$ may exceed the dimension
of the Hilbert space and, therefore, the set of the substates is,
in general, overcomplete. 
As a result, the expansion Eq.(\ref{Rep5}) of  $|\Phi^A[\varphi]\ra$ is non-unique and
also allows for non-trivial (i.e., non-diagonal in the indices
$a(t)$ and $\tilde{a}(t)$) representations of identity).
\newline
In a similar manner, 
we obtain the transformation between the sets of states
corresponding to two finite time measurements $(M=1)$
of the type discussed in Sect.3 of operators
$\hat{A}$ and $\hat{B}$
\begin{equation} \label{Rep2a}
|\Phi^B(f)\ra =\int Df' \hat{U}[f-f']
|\Phi^A(f')\ra ,
\end{equation}
where, explicitly,
\begin{equation} \label{Rep8}
\hat{U}[f-f']\equiv
\int d\lambda \exp[i\lambda(f-f')]
\exp[-i\int_0^T (\hat{H}+\lambda \beta_{B}(t)\hat{B})]
\exp[i\int_0^T(\hat{H}+\lambda \beta_{A}(t)\hat{A})].
\end{equation}
For the amplitudes 
$$ c_{a'}(f')\equiv \la a'|\Phi^A(f')\ra\quad
d_{b'}(f)\equiv \la b'|\Phi^B(f)\ra$$
we have 
\begin{equation} \label{Rep9}
d_{b'}(f)=\sum_{a'} \int df'
\la b'|\hat{U}(f-f')|a'\ra c_{a'}(f')
\end{equation}
which cannot, in general, be simplified further.
\newline
Finally, the transformation between  
the amplitudes, corresponding to two 
von Neumann measurements, 
each taken at the time $t=T$,
of $\hat{A}$ and $\hat{B}$ can be obtained
from Eqs.(\ref{Rep8}) and (\ref{Rep9}) by choosing $ \beta_{A}(t), \beta_{B}(t)\rightarrow \delta(t-T)$.
As a result, the operators in the r.h.s. of Eq.(\ref{Rep8})
factorise, e.g., $\exp[-i\int_0^T (\hat{H}+\lambda \beta_{B}(t)\hat{B})]
\approx \exp[-i\lambda \hat{B}] \exp[-i\hat{H}T]$,
and taking $T\rightarrow 0$ we obtain
\begin{equation} \label{Rep10a}
|\Phi^{Z}(t|f)\ra =\sum_{z}\delta(f-z)|z\ra \la z|\Psi_0\ra \quad Z=A,B.
\end{equation}
and
\begin{equation} \label{Rep10}
\hat{U}[f-f']=\sum_{a,b}\delta(f-f'-b+a)|b\ra \la b|a\ra \la a|.
\end{equation}
Inserting Eqs.(\ref{Rep10a}) and (\ref{Rep10})
into Eq.(\ref{Rep9}) and integrating over $f$ yields
\begin{equation} \label{Rep11}
\la b|\Psi_0\ra =\sum_{a}\la b|a\ra \la a|\Psi_0\ra .
\end{equation}
which is the relation between components of a vector $|\Psi_0\ra$
in the basis sets $\{|a\ra \}$ and $\{|b\ra \}$,
interpreted as the probability amplitudes
to have the values $a$ and $b$ in the state $|\Psi_0\ra$.
This allows to transform amplitudes for finding,
in the state $|\Psi_T\ra$, the values
of the variable $\hat{A}$ into those for finding the values of
$\hat{B}$ \cite{Mess}.
Note that in the von-Neumann case the subsets $|\Phi^{A,B}[\varphi]\ra$ 
form, provided none of the $\la a|\Psi_0\ra$ vanish,
 a complete orthogonal basis, 
in which any given state can be expanded in a unique manner.

\section{Commuting operators and Feynman's functionals.}

We proceed to considering,
first in an
$n$-dimensional Hilbert space,
the class of operators
which commute 
with a given operator
$\hat{A}$, whose eigenvalues,
$a_j, j=1,...n$
are assumed to be
non-degenerate. 
Such operators share with $\hat{A}$ its complete set of eigenstates,
$|a_j\ra , j=1,...n$, and can be written in the form
\begin{equation} \label{Co1}
\hat{A}'= F(\hat{A}) \equiv \sum_{j=1}^n |a_j\ra F(a_j) \la a_j|,
\end{equation}
where the eigenvalues $F(a_j)$, $j=1,2,...$, may or may not 
 be all different.
The fine grained decomposition 
$|\Phi^{F(A)}[\varphi]\ra$ for the operator $F(\hat{A})$ can then be written as
\begin{equation} \label{Co2}
|\Phi^{F(A)}[\varphi]\ra =
\int D\varphi' \delta[\varphi-F(\varphi')]|\Phi^{A}[\varphi]\ra .
\end{equation}
Indeed, the set of substates obtained for the operator $\hat{A}$,
$|\Phi^{A}[\varphi]\ra$ is given by Eq.(\ref{EG3}) and
integration of (\ref{Co2}) over $D \varphi'$ yields the same
form, but with $\hat{A}$ replaced by $F(\hat{A})$.
If none of the eigenvalues of $F(\hat{A})$, $F(a_j)$, are degenerate, 
the two sets of substates are identical,
and there is one-to-one correspondence between
the sets of eigenpaths, i.e., 
the same substate correspond to the eigenpath $\varphi (t')= a(t')$ and 
for the operator $\hat{A}$ and the eigenpath 
$\varphi'(t')=F(a(t'))$, for the operator $F(\hat{A})$.
If, on the other hand, some of the eigenvalues are degenerate,
e.g., $F(a_m)=F(a_n)$, several paths $a(t)$ become indistinguishable 
and cannot be told apart by a measurement
of $F(\hat{A})$, $\varphi (t)\equiv a_m$ and $\varphi' (t)\equiv a_n$ being 
two obvious examples. In the extreme case $F(a_1)=F(a_2)=...=F(a_n)=F_0$,
i.e., $F(\hat{A})=F_0=const$ 
the set of eigenpaths collapses to a single constant 
path $\varphi = F_0$
and the solution of Eq.(\ref{clockf}) takes the form (cf. Eq.(\ref{EG3}))
\begin{equation} \label{Co3}
|\Phi^{F_0}(T|[\varphi])\ra = \delta[\varphi - F_0]|\Psi_T\ra .
\end{equation}
In this case, no meaningful decomposition of the Schr\"odinger state
$\Psi_T$ is obtained and no information abut the system may be gained.
\newline
With the help of Eq.(\ref{Four2la}), the amplitude for a finite
time measurement of $F(\hat{A})$ involving a single
meter (the case of several meters can be analysed in the same way)
becomes
\begin{equation} \label{Co4}
|\Phi^{F(A)}(T|f)\ra = \sum_{[a]}\delta(f-F[a]) |\Phi^{A}(T|[a])\ra .
\end{equation}
This is a restricted path sum in which a particular history
$a(t)$ does or does not contribute to the substate
$|\Phi^{F(A)}(T|f)\ra$ depending on whether the value of the 
{\it functional}
\begin{equation} \label{Co5}
F[a]\equiv \int_0^T \beta(t') F(a(t')) dt'
\end{equation}
is or is not equal to $f$. 
The functionals defined on the Feynman paths,
$a(t')=x(t')$,
were introduced in Ref. \cite{Feyn} and
are worth a brief discussion.
Applying Eqs.(\ref{Co1}) to (\ref{Co5}) to the Feynman path integral
(\ref{EG9}) allows one to analyse any observable which commutes with the particle's
coordinate $x$, $F(x)$.
With the particular choice $\beta(t)=T^{-1}=const$,
the substate $|\Phi^{F(x)}(T|f)\ra$ becomes 
the result of evolution of the initial state
$|\Psi_0\ra$ along those and only those Feynman paths, for which
 the time average of $F(x)$, 
\begin{equation} \label{Co6}
\la F(x)\ra _T \equiv T^{-1} \int_0^T F(x(t)) dt
\end{equation}
equals $f$. The scalar product $\la x|\Phi^{F(x)}(T|f)\ra$ yields
the amplitude for a particle at $t=T$ at a location $x$
to have a definite value $f$ of $\la F(x)\ra _T$ in the past
and the Schr\"odinger amplitude $\la x|\Psi_T\ra$ can be seen as a result
of interference between different mean values of the variable $F$. 
It follows from Eqs.(\ref{phil}) and (\ref{EG9}), the amplitudes $\la x|\Phi^{F(x)}(T|f)\ra$ can
be found by solving the Schr\"odinger equation corresponding
to the modified classical action 
$$S_{\lambda}[x(t)]=S[x(t)]+\lambda \int_0^T F(x)dt$$
containing and extra potential,
$-\lambda F(x)$, and then taking the Fourier transform with 
respect to $\lambda$.
The possibility of using Eq.(\ref{Co6}) as a starting point 
for formulating  measurement theory in the coordinate space
has been studied in Refs.\cite{S1,S2,S3,S4,S5,S6,S7,S8,S9,S10,S11,
S12,S13,S14,SBOOK}.
The case of the mean coordinate, $F(x)=x$, was analised in 
\cite{S1}. The choice $F(x)=\theta_{\Omega}(x)$ was used
to define the quantum traversal tine and extensively studied 
in \cite{SBOOK}.
The same technique was used in \cite{S13}
in order to analyse the amount of time  a qubit spends in given 
quantum state. 

\section {Simultaneous histories for non-commutimg observables. 
The phase space path integral}

Until now we have analysed the histories generated by a 
single variable $\hat{A}$. Next we consider the possibility
of constructing histories containing simultanious information
about two non-commuting observables $\hat{A}$ and $\hat{B}$.
In the following we will assume that the commutator of a
$\hat{A}$ and $\hat{B}$ is an imaginary $c$-mumber
\begin{equation} \label{AB1}
[\hat{A},\hat{B}] = 2iC,
\end{equation}
which is the case, for example, for the 
canonically conjugate momentum and coordinate
operators, $\hat{p}$ and $\hat{q}$.
Accordingly, we modify Eq.(\ref{rec}) ($\bar {\varphi}=\{\varphi_1,\varphi_2\}$)
\begin{equation} \label{recab}
i\partial_{t}|\Phi(t|[{\bar {\varphi}}])\ra =\{\hat{H}-i\hat{A}\frac{\delta}{\delta
\varphi_1(t)}-i\hat{B}\frac{\delta}{\delta
\varphi_2(t)}\}|\Phi(t|[\bar {\varphi}])\ra ,
\end{equation}
and impose the initial condition
\begin{equation} \label{AB3}
|\Phi(t=0|[\bar {\varphi}])\ra =|\Psi_0\ra\delta[\varphi_1]\delta[\varphi_2].
\end{equation}
As is Sect.2, the (now two-dimensional) Fourier transform $|\Phi(t|[\bar {\lambda}])\ra$ 
satisfies a time-dependent Schr\"odinger equation and 
can be written (cf. Eq.(\ref{F1}))
\begin{equation} \label{AB4}
|\Phi(t|[\bar{ \lambda}])\ra =\exp[-i\int_0^t(\hat{H}+\lambda_1(t')\hat{A}+
\lambda_2(t')\hat{B})dt']|\Psi_0\ra .
\end{equation}
Slicing the time interval into $N$ segments of length $\epsilon$,
and applying the Trotter  and the Baker-Hausdorff 
formulae \cite{Shul} to factorise
$\exp(-i\hat{H}\epsilon)$, $\exp(-i\lambda_1\hat{A}\epsilon)$ and
$\exp(-i\lambda_2\hat{B}\epsilon)$ we obtain
\begin{eqnarray} \label{AB5}
\exp[-i\int_0^t(\hat{H}+\lambda_1(t')\hat{A}+
\lambda_2(t')\hat{B})dt']|\Psi_0\ra \approx \prod_{j=1}^N 
\exp(-i\hat{H}\epsilon)\times  \\
\exp(-i\lambda_1\hat{A}\epsilon)\times
\exp(-i\lambda_2\hat{B}\epsilon)\times \exp(-iC\lambda_1 \lambda_2 \epsilon^2).
\end{eqnarray}
where we have retained the term containing the commutator
$2iC$ even though it is quadratic in $\epsilon = T/N$.
Performing the inverse Fourier transform and taking into account 
the convolution property (\ref{ACON}), we obtain
\begin{equation} \label{AB6}
|\Phi(T|[\bar {\varphi}])\ra
=\int D\bar {\varphi }' u[\bar {\varphi }-\bar {\varphi }']
|\Phi'(T|[\bar {\varphi }])\ra ,
\end{equation}
where 
\begin{equation} \label{AB6a}
|\Phi'(T|[\bar {\varphi}])\ra\equiv \sum_{[a,b]} \delta[\varphi_1-a]
\delta[\varphi_1-b] \hat{U}_T[a,b],
\end{equation}
and
\begin{equation} \label{AB6ab}
\hat{U}_T[a,b] \equiv lim_{N \rightarrow \infty}\prod_{j=1}^{N}
\exp(-i\hat{H}\epsilon)|b_{k_j}\ra\la b_{k_j}|a_{k_j}\ra\la a_{k_j}|.
\end{equation}
We note that $|\Phi'\ra$ is constructed from two-dimenisonal eigenpaths,
in which both $A$ and $B$ have well defined values at any time $t'$,
in a way similar to that the fine grained substates were constructed
for a single variable $A$ in Sect.5.
The substates $|\Phi\ra$, corresponding to the initial condition
(\ref{AB3}) are connected to $|\Phi'\ra$ by a unitary transformation
with the kernel
\begin{equation} \label{AB7}
u[\bar {\varphi}]\equiv lim_{N \rightarrow \infty}
(2\pi/c)^N\prod_{j=1}^N \exp[i\varphi_1(t_j)\varphi_2(t_j)/C],
\end{equation}
which indicates that the values of two non-commuting variables
cannot have well defined values at the same time.

For a quantum particle of mass $m$ in one-dimensional potential $V(q)$,
and
$\hat{A}\equiv \hat{p}$, $\hat{B}\equiv \hat{q}$, $C=1$,
$\hat{U}[p,q]$ can, using 

\begin{equation} \label{PQ4}
\la p|q\ra =\la q|p\ra ^{*}=\exp[ipq],
\end{equation}
be written as \cite{Shul,Klein}
\begin{equation} \label{AB8a}
\hat{U}[a,b] =
\int dq_T dq_0 Dp Dq |q_T\ra \exp\{i\int_0^T [p\dot{q}-H(p,q)] dt\}\la q_0|
\end{equation}
where $\int_0^T [p\dot{q}-H(p,q)] dt \equiv \lim_{N\rightarrow \infty}
\sum_{j=1}^N \epsilon [p_j(q_j-q_{j-1})/\epsilon -p_j^2/2m-V(q_j)i]$.
It is easy to check that the unitary kernel $u[\bar{\varphi}-\bar{\varphi}']£$,
which arises from the exponential of the commutator in Eq.
(\ref{AB5}),
has the property
\begin{equation} \label{AB9}
\int D\bar {\varphi}_{1,2} ' u^{*}[\bar {\varphi}-\bar {\varphi}']
= \delta[\varphi_{2,1}],
\end{equation}
so that integrating Eq.(\ref{AB5}) over $D\varphi_1 D\varphi_2$
yields the standard phase integral representation for the 
Feynman propagator \cite{Shul,Klein},
\begin{equation} \label{AB10}
\la q_T|\Psi_T\ra = \int  dq_0 Dp Dq |q_T\ra \exp\{i\int_0^T [p\dot{q}-H(p,q)] dt\}
\la q_0|\Psi_0\ra .
\end{equation}
\newline
Note that, in general, the decomposition 
$|\Phi^{F(A),F'(B)}(T|[\bar {\varphi}])\ra$ for the two operators 
$\hat{A'}$ and $\hat{B'}$
\begin{equation} \label{AB10a}
\hat{A'}=F(\hat{A}), \quad \hat{B'}=F(\hat{B}),
\end{equation}
cannot be obtained from $|\Phi(T|[\bar {\varphi}])\ra$ 
in a way it was done in Sect.8 for a single operator, 
$F(\hat{A})$,
because, in general the commutator of $F(\hat{A})$ and 
$F(\hat{B})$ is not
a $c$-number and the derivation leading to Eq.(\ref{AB6})
no longer applies.
\newline
To conclude, we 
leave aside an interesting question
of simultaneous measurement of non-commuting variables \cite{AK}
and briefly discuss the possibility of constructing, with the help
of our detailed knowledge of $|\Phi(T|[\bar {\varphi}])\ra$,
histories for an operator function, 
$F(\hat{A},\hat{B})$, of the non-commuting 
variables $\hat{A}$ and $\hat{B}$. We shall limit ourselves
to the simplest choice 
\begin{equation} \label{AB12}
F(\hat{A},\hat{B})=\hat{A}+\hat{B}.
\end{equation}
We note first that for any functional $G[\varphi-(\varphi_1+
\varphi_2)]$ and any solution $|\Phi(T|[\bar {\varphi}])\ra$
of Eq.(\ref{AB1}),
\begin{equation} \label{AB12a}
|\Psi(t|[\varphi])\ra\equiv \int D\varphi_1 D\varphi_2
G[\varphi-(\varphi_1+
\varphi_2)] |\Phi(t|[\bar {\varphi}])\ra
\end{equation}
satisfies the 'recorder' equation (\ref{rec}) with 
$\hat{A}$ replaced by $\hat{A}+\hat{B}$. As in Sect.2,
the proof is obtained by multiplying Eq.(\ref{AB1}) by
$G$ on the left and integrating by parts taking into
account that $\delta G/\delta \varphi_{1,2} = -\delta G/\delta \varphi$.
Putting $G=\delta[\varphi-(\varphi_1+
\varphi_2)]$ and choosing $|\Phi\ra$ in Eq.(\ref{AB12}) yields
the solution $|\Phi_{A+B}(t|[\varphi])\ra$ with the initial
condition 
\begin{equation} \label{AB13}
|\Phi^{A+B}(t=0|[\varphi])\ra =|\Psi_0\ra
\int D\varphi_1 D\varphi_2 \delta[\varphi-(\varphi_1+
\varphi_2)]\delta[\varphi_1]\delta[\varphi_2]=
|\Psi_0\ra \delta[\varphi].
\end{equation}
This is the fine-grained decomposition for the single variable
$\hat{A}+\hat{B}$ as discussed in Sect.2.
Evaluating the integral in the r.h.s. of Eq.(\ref{AB12})
with the help of Eq.(\ref{C5}) and performing the Gaussian integrations yields
\begin{equation} \label{AB15}
|\Phi^{A+B}[\varphi]\ra =
\sum_{[a,b]} \tilde{u}[\varphi-(a+b)] 
\hat{U}[a,b]|\Psi_0\ra
\end{equation}
where
\begin{eqnarray} \label{AB16}
\tilde{u}[\varphi]\equiv 
\int D\varphi_1 D\varphi_2 \delta[\varphi-(\varphi_1+
\varphi_2)] u[\varphi_1,\varphi_2]= \\ \nonumber
lim_{N \rightarrow \infty}
(2\pi/C)^N\prod_{j=1}^N \exp[i\varphi^2(t_j)/4C].
\end{eqnarray}
We see, therefore, that since for two non-commuting operators
the value of $\hat{A}+\hat{B}$ is not equal to the sum
of those of $\hat{A}$ and $\hat{B}$, 
we cannot assign sharply defined values of
$a(t')+b(t')$ to a path $\varphi$.
The uncertainty inherent in such an assignment is
determined by the value of the commutator $2iC^{1/2}$
of the two observables.
Finally, as shown in the Appendix E, for a finite time measurement of $A+B$ 
contribution from the term, containing the commutator vanishes
and the fine grade measurement amplitude may be written the restricted 
eigenpath sum 
\begin{equation} \label{AB17}
|\Phi'(T|f)\ra\equiv \sum_{[a,b]} 
\delta \Big(f-\int_0^T \beta(t') a(t') b(t') dt'\Big) \hat{U}_T[a,b]|\Psi_0\ra ,
\end{equation}
where, for simplicity, we considered one meter only ($M=1$).

\section {Conclusions}

For a variable of interest, we have introduced virtual 
paths, or histories, such that various measurement amplitudes can 
be obtained as restricted path sums.
Our analysis of quantum measurement on an single quantum
system is similar to that of a double slit experiment,
in that a measurement implies replacing a coherent superposition
of certain 'routes' leading to the current state of a system,
by one in which the routes become, at the cost of destroying
interference effects, exclusive or nearly exclusive alternatives.

We conclude with a more detailed summary.
For a given variable $A$, we define a path (history)
as all its values, $\varphi (t)$, specified within a time
interval $0\le t' \le T$.
At the time $T$, each such history contributes a substate
$|\Phi [\varphi]\ra$ to the Schr\"odinger state of the system
$|\Psi_T\ra$. The decomposition $|\Phi [\varphi]\ra$ contains the most
detailed information about the past value of $A$
and can be obtained by evolving the system's initial state
in accordance with the 'recorder' equation (\ref{rec}),
which assigns substates to each $\varphi(t')$.
Further use of the substates depends on the 
the physical conditions imposed on the system.
For a system in isolation, the substates add up coherently
to produce a pure state $|\Psi_T\ra$,
and all information about the values of $A$ is lost through
interference, just like there is no knowledge of the 
path taken by an electron or a photon in a double slit or gravitational
lensing experiment if an interference pattern is observed. 
Bringing the system into contact with a meter, meters or a measuring
environment, destroys coherence between the substates, and the system
ends up in a mixed, rather than pure, state. It is the defining property
of a meter, designed to measure $A$,
that it distinguishes between  classes of paths, typically labelled
by the value $f$ of a functional or functionals.
Within each class, the substates add up coherently,
so that only part of all information contained in $|\Phi [\varphi]\ra$
is extracted, and the meter's reading does not determine the past path
uniquely.
For a realistic meter, the initial pointer position is always
somewhat uncertain.
As a result, for each class of paths, the value of $f$
is not sharply defined but rather has an error margin $\Delta f$.
Finding at $t=T$ a meter's reading $f$ indicates that the system has evolved
along the paths for which the value of the functional effectively
lies between $f-\Delta f$ and  $f+\Delta f$.
In a number of identical trials, this occurs with the probability
$\la \Psi(f)|\Psi(f)\ra$, where $|\Psi(f)\ra$ is the coherent sum of
all substates consistent with the reading.
A more accurate meter perturbs the system to a  larger extent.
Different choices of number of meters, their accuracies and 
durations over which each of them interacts with the system,
provide various ways to measure the variable $A$.

Non-zero substates 
can be attributed only to the eigenpaths, i.e., the paths 
such that at each moment in time,
$\varphi(t')$ coincides with one of the eigenvalues 
of the operator $\hat{A}$, $a_i$.
For an eigenpath, the substate proportional is to the 
eigenstate of $\hat{A}$ corresponding to $\varphi(T)$, 
with the coefficient dependent on the path,
so that, in general, the number of substates exceeds 
the dimension of the Hilbert space.
A system evolves, as one would 
expect, along the virtual paths
which cannot live in its Hilbert space.
For a structureless  particle in one dimension,
the eigenpaths of the position operator  are the Feynman paths,
and the sum over such paths yields the Feynman path integral.
For a qubit,
 the sum over virtual eigenpaths, which alternate between
the two eigenvalues of $\hat{A}$ yields the perturbation expansion
for the system's state in the representation which diagonalises
$\hat{A}$.

Two non-commuting variables, $\hat{A}$ and $\hat{B}$,
produce two different sets of eigenpaths and substates, which can be 
expressed in terms of each other.
Because the sets are, in general overcomplete,
such expansion is not unique.
An exception is a von Neumann-like measurement which
determines the instantaneous value of a variable at the time 
of measurements. For such a measurements, the substates form an orthonormal
sets connected by unitary transformations. 
For two commuting variables with non-degenerate eigenvalues,
the two sets of eigenpaths are in one-to-one correspondence and
the sets of substates are identical.
If some of the eigenvalues of, say $B$, are degenerate,
then some of the substates generated by $B$ are coherent sums
of those corresponding to $A$. In this case,
a measurement amplitude for the variable $B$ can be obtained
as a restricted sum over the paths obtained for the 
non-degenerate variable $A$.
For a structureless particle in the coordinate space
this allows to analyse the measurements of various functions
of the particle's coordinate $x$ in terms of the Feynman paths,
as was done in Refs. \cite{SBOOK} for the quantum traversal time.

Simultaneous histories can be constructed for
two (or more)  non-commuting variables by decomposing the
Schr\"odinger state into substates labelled by a two-component
path index $\{\varphi_1(t'),\varphi_2(t')\}$.
For two canonically conjugate variables, as the example in Sect.9 shows,
the subsets can still be expressed in terms
of 'simultaneous eigenpaths' mixed with the unitary kernel,
containing the non-zero commutator of $\hat{p}$ and $\hat{q}$.
Summing such substates over all possible paths yields the
phase space representation for the Schr\"odinger state 
of the system.
The present approach will be used in
future work in order to address such issues as 
weak measurements,
measurements conducted on composite systems, more detailed analysis
of non-commuting observables  and possible generalisations
of 'recorder' equation of Sect.2  which have fallen outside the scope of this paper.

\section {Appendix A: The functional Fourier transform.}

Consider a set $\{\varphi(t)\}$ of all continuous, but not 
necessarily smooth real functions defined on an interval $0\le t \le T$
with arbitrary boundary values $\varphi(0)$ and $\varphi(T)$.
Slicing the interval into $N$ subintervals of the length $\epsilon
= T/N$ we define a functional  $F[\varphi]$ as the limit 
\begin{equation} \label{A1}
F[\varphi] = lim_{N\rightarrow \infty} F(\vec{\varphi})
\end{equation}
where $\vec {\varphi}\equiv (\varphi_1,\varphi_2,...\varphi_{N+1})$,
 $\varphi_i=\varphi(\epsilon(i-1))$ and  $F(\vec{\varphi})$ is a known
function for each value of $N$. For example, a functional 
\begin{equation} \label{A2}
I[\varphi]\equiv \int_0^T \varphi(t) \lambda (t) dt =
lim_{N\rightarrow \infty} \sum_{i=1}^{N+1} \varphi_i \lambda_i \epsilon
\end{equation}
is defined by its discretised Riemann sum.
Further, the functional derivative is defined as 
\begin{equation} \label{A3}
\delta F[\varphi]/\delta \varphi (t) = lim_{N\rightarrow \infty}
\epsilon ^{-1} \partial F(\vec{\varphi})/\partial \varphi_m,
\quad m=t/\epsilon ,
\end{equation}
so that for $I[\varphi]$ in Eq.(\ref{A2}) $\delta I[\varphi]/\delta \varphi (t)
= \lambda (t)$, as it should.
For the sum over the functions $\varphi$ we have
\begin{equation} \label{A4}
\int_{\{\varphi\}} D\varphi F[\varphi] \equiv
lim_{N\rightarrow \infty} \int_{-\infty}^{\infty} d \varphi_1 d\varphi_2
...d\varphi_{N+1} F(\vec{\varphi})
\end{equation}
It is readily seen that if $\lim_{\varphi(t)\rightarrow \pm \infty}
F[\varphi] = 0$, then
\begin{equation} \label{A5}
\int D\varphi \delta F[\varphi]/\delta \varphi (t) =0.
\end{equation}
If we define the functional Fourier transform for $F[\varphi]$
as
\begin{equation} \label{A6}
\tilde{F}[\lambda ] \equiv lim_{N\rightarrow \infty} 
(\epsilon / 2\pi )^{N+1} \tilde{F}(\epsilon \vec{\lambda})
\end{equation}
(where $\tilde{F}(\vec{\lambda})\equiv 
\int F(\vec{\varphi}) \exp(-i\vec{\lambda}.
\vec{\varphi})$ and $\vec{\lambda}.
\vec{\varphi} \equiv \sum_{j=1}^{N+1} \lambda_j \varphi_j$),
$F[\varphi]$ can be written as a Fourier functional integral
\begin{equation} \label{A7}
F[\varphi] = \int D\lambda \tilde{F}[\lambda ] \exp [i\int_0^T
\lambda (t) \varphi (t) dt].
\end{equation}
A particular choice 
\begin{equation} \label{A8}
\tilde{F}[\lambda ] \equiv \tilde{\delta}[\lambda] =lim_{N\rightarrow \infty}
(\epsilon / 2\pi )^{N+1} 
\end{equation}
yields the $\delta$ functional ($\delta(z)$ is the Dirac 
$\delta$-function)
\begin{equation} \label{A9}
\delta [\varphi] = lim_{N \rightarrow \infty} \prod_{j=1}^{N+1}
\delta (\varphi_j)
\end{equation}
with the obvious property
\begin{equation} \label{A10}
\int D\varphi F[\varphi] \delta [\varphi] = F[\varphi \equiv 0].
\end{equation}
We will also require the convolution property
\begin{equation} \label{ACON}
\int D\lambda \tilde{F}[\lambda ] \tilde{G}[\lambda ]\exp [i\int_0^T
\lambda (t) \varphi (t) dt]
=\int D\varphi' F[\varphi-\varphi'] G[\varphi],
\end{equation}
which can be obtained by considering the time discretised Fourier transform.

\section {Appendix B: Lie-Trotter and Baker-Campbell-Hausdorff formulae}

The generalised Lie-Trotter formula reads \cite{TROT}
\begin{equation} \label{T1}
\lim_{N\rightarrow \infty}[\hat{F}(t/N)]^N = \exp[t\partial_t \hat{F}(0)]
\end{equation}
where $\hat{F}(t)$ is an operator function of $t$ such that
\begin{equation} \label{T2}
\hat{F}(0)=1.
\end{equation}
Choosing
$$\hat{F}(t)=\exp[t\hat{A}] \exp[t\hat{B}]$$
yields the Trotter product formula
\begin{equation} \label{T3}
\lim_{N\rightarrow \infty} \{\exp[t/N\hat{A}] \exp[t/N\hat{B}]\}^N = \exp[t(\hat{a}+\hat{b})].
\end{equation}

The Baker-Campbell-Haussdorff identity states that for two
operators $\hat{A}$ and $\hat{B}$, such that
\begin{equation} \label{T4}
[\hat{A},[\hat{A},\hat{B}]]=[\hat{B},[\hat{B},\hat{A}]]
\end{equation}
where the square brackets denote the commutator,
\begin{equation} \label{T5}
\exp[\hat{A}+\hat{B}]=\exp[\hat{A}]\exp[\hat{B}]\exp\{-[\hat{A},\hat{B}]/2\}
\end{equation}

\section {Appendix C: The eigenpath expansion as a solution of the
'recorder' equation.}

In order to verify that the eigenpath expansion (\ref{EG3}) satisfies 
Eq.(\ref{rec}),
consider a more general form

\begin{equation} \label{B1}
|\Psi(t|[\varphi])\ra=\sum_{[a]}G[\varphi-u_t a]|\Phi_t[a]\ra
\end{equation}

where $G[\varphi]$ is an arbitrary functional, $u_t(t')$
is a function of $t'$, which also depends on the time $t$
and $|\Phi_t[a]\ra$ is defined in Eq.(\ref{EG3a}).
Then 
 
\begin{eqnarray} \label{B2}
\partial_t |\Psi(t|[\varphi])\ra=
-\sum_{[a]}\int_0^T \frac {\delta G}{\delta \varphi (t')}
\partial_t u_t(t')a(t')dt'|\Phi_t[a]\ra
+\sum_{[a]}G[\varphi-u_t a]\partial_t|\Phi_t[a]\ra.
\end{eqnarray}

Using Eq.(\ref{EG4}) we have

\begin{equation} \label{B3}
\partial_t|\Phi_t[a]\ra =-i\hat{H}|\Phi_t[a]\ra ,
\end{equation}

and choosing

\begin{equation} \label{B4}
u_t=\theta_t(t'), \quad \partial_t u_t =\delta(t-t')
\end{equation}

yields

\begin{eqnarray} \label{B5}
\partial_t |\Psi(t|[\varphi])\ra =
-\frac{\delta}{\delta \varphi(t)}\sum_{[a]}G[\varphi-\theta _t a]a(t)
|\Phi_t[a]\ra
-i\hat{H}\sum_{[a]}G[\varphi-\theta _t a]|\Phi_t[a]\ra .
\end{eqnarray}

For $G[\varphi]=\delta[\varphi]$, with the help of the relation

\begin{equation} \label{B6}
\hat{A}|\Phi_t[a]\ra =a(t)|\Phi_t[a]\ra ,
\end{equation}

Eq.(\ref{B5}) reduces to Eq.(\ref{rec}).

\section {Appendix D: Finite time measurements and the Mensky's
formula.}

We will show that in Eq.(\ref{Four2la}) for the set $|\Phi(t|\vec{f})\ra$,
$|\Phi(t,|[\varphi])\ra$ can be replaced by $|\Phi_W(t,|[\varphi])\ra$
defined in Eq.(\ref{D9a}) of Sect.6.  From Eq.(\ref{ACON}) we have
\begin{equation} \label{DD1}
|\Phi_W(t,|[\lambda])\ra=G_W[\lambda]|\Phi(t,|[\lambda])\ra ,
\end{equation}
where
\begin{equation} \label{DD2}
G_W[\lambda]=lim_{N\rightarrow \infty}C\prod_{j=1}^N\exp[-\lambda_j^2 \sigma^2 \epsilon /4]
=C'\exp[-\sigma^2\int_0^T \lambda ^2 dt'/4].
\end{equation}
Using Eq.(\ref{C5}) for the Fourier transform of the finite time
measurement set computed with the help of $|\Phi_W(t,|[\varphi])\ra$
we have ($M=1$)
\begin{equation} \label{DD3}
|\Phi_W(t,|\lambda)\ra\equiv lim_{\sigma \rightarrow 0}\exp[-\lambda^2\sigma^2\int_0^T \beta ^2 dt'/4]
|\Phi(t,|[\lambda \beta])\ra .
\end{equation}
Therefore for any $\beta(t)$ such that $\int_0^T \beta ^2 dt'< \infty$ the first
factor in Eq.(\ref{D3}) can be replaced by unity, and, therefore,
$|\Phi(t,|\lambda)\ra$ can be used in place of $|\Phi_W(t,|\lambda)\ra$.

\section {Appendix E: Some properties of restricted path sums.}

Inserting the relations (we write $|\Phi[\varphi]\ra$ and 
$|\Phi(\vec{f})\ra$ for $|\Phi(t,|[\varphi])\ra$ and 
$\Phi(t|\vec{f})\ra$, respectively )

\begin{equation} \label{CC1}
|\Phi[\varphi]\ra =\int D\lambda \exp(i\int \lambda \varphi dt')
|\Phi[\lambda]\ra ,
\end{equation}

\begin{equation} \label{CC2}
|\Phi(\vec{f})\ra =\int d\vec{\lambda} \exp(i \vec{\lambda}\vec{f})
|\Phi (\vec{\lambda})\ra
\end{equation}
and
\begin{equation} \label{CC3}
\delta(f_i-\int_0^T \beta_i \varphi dt')
= (2\pi)^{-1}
\int d\lambda_i \exp[i \lambda_i (f_i-\int_0^T \beta_i \varphi dt')]
\end{equation}
into the definition
\begin{equation} \label{CC4}
|\Phi(\vec{f})\ra \equiv
\int D\varphi \prod_{i=1}^M \delta(f_i-\int_0^T \beta_i \varphi dt')
|\Phi[\varphi]\ra
\end{equation}
yields a simple relation between the Fourier transforms of the
measurement amplitude $|\Phi(\vec{f})\ra$ and the fine grained
set $|\Phi[\varphi]\ra$,

\begin{equation} \label{CC5}
|\Phi(\vec{\lambda})\ra =|\Phi[\lambda]\ra |_{\lambda =\sum_{i=1}^M\lambda_i\beta_i}.
\end{equation}

Further, for the fine grained set $|\Phi^{A+B}[\varphi]\ra$ in Eq.(\ref{AB15}),

\begin{equation} \label{CC6}
|\Phi^{A+B}[\varphi]\ra\equiv
\int D\varphi_1 D\varphi_1 \delta [\varphi-\varphi_1-\varphi_2]|\Phi^{A,B}[\varphi_1,
\varphi_2]\ra
\end{equation}

we find

\begin{equation} \label{CC7}
|\Phi^{A+B}[\lambda]\ra =
|\Phi^{A,B}[\lambda,\lambda]\ra
\end{equation}

where $|\Phi^{A,B}[\lambda_1,\lambda_2]\ra$ is the Fourier transform
of $|\Phi^{A,B}[\varphi_1,\varphi_2]\ra$ in Eq.(\ref{recab}).
Combining Eqs.(\ref{C5}) and (\ref{C7}) yields

\begin{equation} \label{CC8}
|\Phi^{A+B}(\vec{\lambda})\ra=|\Phi^{A,B}[\lambda,\lambda]\ra
|_{\lambda =\sum_{i=1}^M\lambda_i\beta_i}.
\end{equation}

\newpage

%\newpage
%\vspace {5cm}
%\begin{figure}[ht]
%\epsfxsize=18cm
%\centering\leavevmode\epsfbox{FIG2.eps}
%\vspace{0pt}
%\caption{ }
%\label{fig:FIG}
%\end{figure}

%\newpage
%\begin{center}
%{\bf Figure captions}
%\end{center}
%{\bf Fig.1: }
%a)
%A schematic diagram of a path $\varphi (t')$
%which contributes to $|\Phi(t|[\varphi])>$
%at $t < T$ (solid). Also shown by a dashed line
%is the tube which contains the paths, contributing
%to $|\Psi(t|[\varphi])>$, obtained by Gaussian coarse
%graining with the width $\sigma$.
%
%b) An eigenpath which contributes to $|\Phi(T|[\varphi])>$
%for a two-level system ( $a_1 = 1$, $a_2=2$)

\newpage
$ $

\vspace{3.0cm}
$ $
\newpage
$ $
\vspace{2.0cm}
$ $
\newpage
\begin{figure}
\includegraphics[width=19cm]{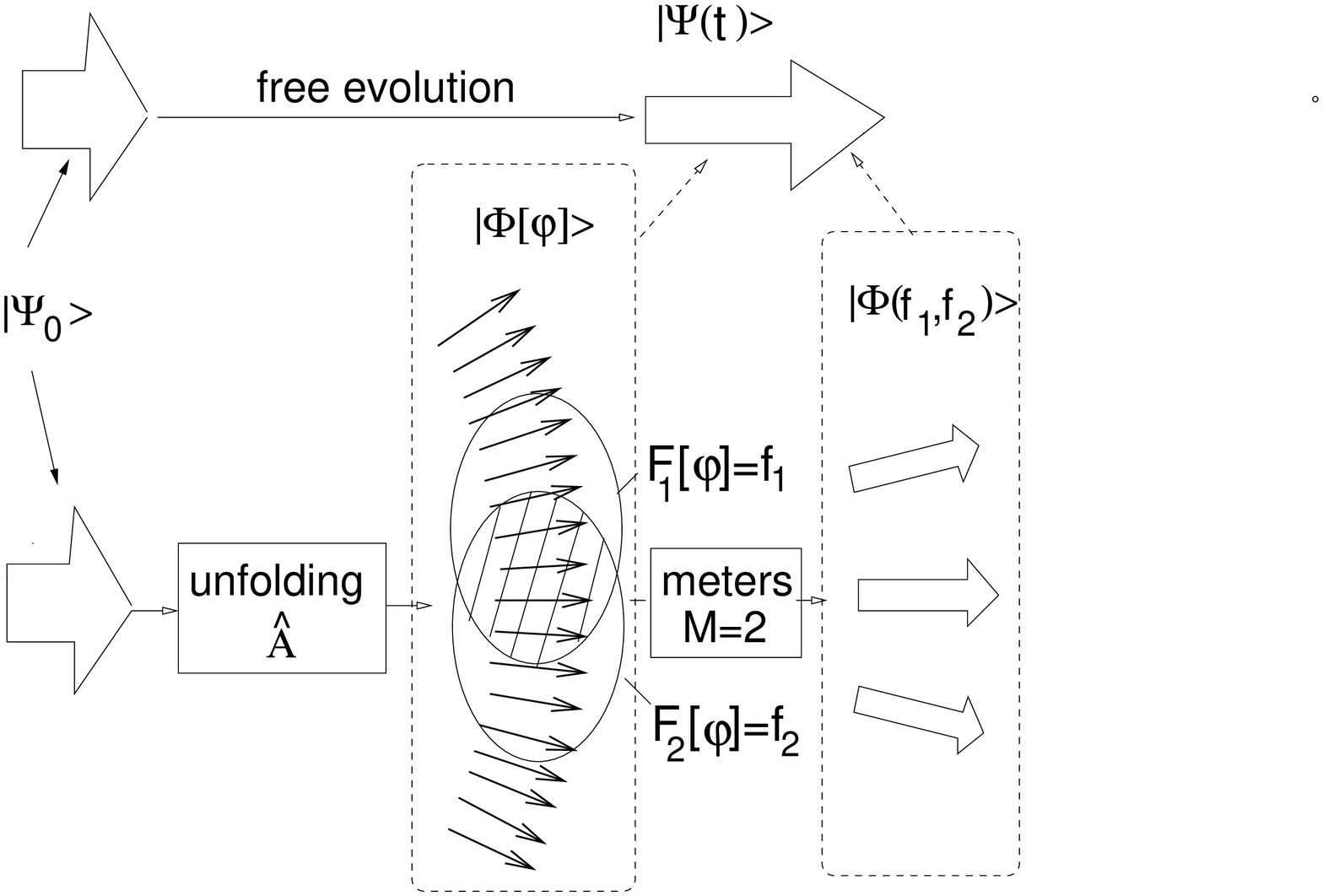}
\vspace{5.0cm}
\caption[]{
A system starts in a state $\Psi_0\ra$ which,
without measurements,
would evolve into $\Psi(t)\ra$.
$\Psi(t)\ra$ can be decomposed into a fine
set of substates $|\Psi[\varphi]\ra$, each labelled by a particular
history $\varphi(t')$.
A meter decomposes $\Psi(t)\ra$ into a less
informative substates, labelled by the value $f$
of functional $F[\varphi]$, which are obtained
by summing $|\Psi[\varphi]>$ subject to restriction
$F[\varphi]=f$.
For the two meters shown, the substates in the shaded
area are such that $F_1[\varphi]=f_1$ and $F_2[\varphi]=f_2$,
and add up coherently to the state $|\Phi(f_1.f_2)\ra$,
whose norm determines the probability to register
both $f_1$ and $f_2$.
}
\label{fig:FIG2}
\end{figure}

%INDEX%%%%%%%%%%%%%%%%%%%%%%%%%%%%%%%%%%%%%%%%%%%%%%%%%%%%%%%%%%%%%%%
% Please check with the editor of your book whether he plans to
% include a "mutual" subject index - if so, please code your entries
% in the standard syntax. For your own purposes you may print your
% "personal" index by using the following commands:
%
%\clearpage
%\addcontentsline{toc}{section}{Index}
%\flushbottom
%\printindex
%%%%%%%%%%%%%%%%%%%%%%%%%%%%%%%%%%%%%%%%%%%%%%%%%%%%%%%%%%%%%%%%%%%%%


\begin{thebibliography}{8.}
\bibitem{FOOTC}
'Coarse graining' may not be the most appropriate term as
it usually implies separation of a set into a number of
well defined non-overlapping subsets. In our case, the subsets
of paths are fuzzy and overlapping. Yet, in the absence of
a better word and to preserve the continuity with the previous
work, we will continue to use it in this paper.
\end{thebibliography}
\end{document}